\documentclass[epj]{svjour} 

\usepackage{graphicx}
\usepackage{subfigure}
\usepackage{epsfig}
\usepackage{amssymb}
\usepackage{graphics}

\newcommand{\sqrtsnn}{\sqrt{s_{_{NN}}}}
\def\mean#1{\ensuremath{\left<#1\right>}}

\begin{document}

\title{High $p_T$ leading hadron suppression in nuclear collisions
at $\sqrtsnn\approx$ 20 -- 200 GeV: data versus parton energy loss models}
\author{David d'Enterria\inst{1}
}                     
\institute{$^1$Nevis Laboratories, Columbia University\\ 
Irvington, NY 10533, and New York, NY 10027, USA}

\date{Received: date / Revised version: date}

\abstract{
Experimental results on high transverse momentum (leading) hadron spectra 
in nucleus-nucleus collisions in the range $\sqrtsnn\approx$ 20 -- 200 GeV 
are reviewed with an emphasis on the observed suppression compared to free space 
production in proton-proton collisions at the corresponding center-of-mass energies. 
The transverse-momentum and collision-energy (but seemingly not the in-medium path length) 
dependence of the experimental suppression factors measured in central collisions 
is consistent with the expectations of final-state non-Abelian parton energy loss in a dense QCD medium. 
\PACS{
     {12.38.Mh}{}  \and
     {13.87.Fh}{}  \and
     {24.85.+p}{}  \and
     {25.75.-q}{} 
      } 
} 
\titlerunning{High $p_T$ leading hadron suppression in nuclear collisions at 
$\sqrtsnn\approx$ 20 -- 200 GeV}
\maketitle

\section{Introduction}
\label{sec:intro}
High-energy nucleus-nucleus collisions offer the only experimental means
known so far to concentrate a significant amount of energy ($\mathcal{O}$(TeV) 
at the RHIC collider) in a ``large'' volume, $\mathcal{O}($10$^3$ fm$^3$), 
under laboratory conditions. The expectations, based on first-principles 
lattice-QCD calculations~\cite{karsch}, are that for energy densities above 
$\varepsilon\approx$ 0.7 GeV/fm$^3$, hadronic matter will undergo a phase transition 
towards an extended volume of deconfined and massless quarks, and gluons: 
the Quark Gluon Plasma (QGP). The scrutiny of this new state of matter aspires 
to shed light on some of the open key questions of the strong interaction 
(confinement, chiral symmetry breaking, structure of the QCD vacuum, 
hadronization) that still evade a thorough theoretical description~\cite{millenium_prizes} 
due to their highly non-perturbative nature.

The production of an extremely hot and dense partonic system in 
relativistic heavy-ion reactions should manifest itself in a variety 
of experimental signatures. One of the first proposed ``smoking guns'' 
of QGP formation was ``jet quenching''~\cite{bjorken82}. Namely, the 
disappearance of the collimated spray of hadrons resulting from 
the fragmentation of a hard scattered parton due to the ``absorption'' 
of the parent quark or gluon as it traverses the dense strongly 
interacting medium produced in the reaction. 
Extensive theoretical work on high-energy parton propagation in a QCD 
medium~\cite{gyulassy90,bdmps,glv,wiedemann} has shown that the main 
mechanism of parton attenuation is of radiative nature: the traversing 
parton loses energy mainly by multiple gluon emission (``gluonstrahlung''). 
This medium-induced non-Abelian energy loss results in several observable 
experimental consequences:
\begin{description}
\item (i) depleted production of high $p_T$ (leading) hadrons~\cite{gyulassy90}, 
\item(ii) unbalanced back-to-back di-jet azimuthal correlations~\cite{appel86,blaizot_mclerran86},
\item(iii) modified parton fragmentation functions (energy flow and particle
multiplicity within the final jets)~\cite{armesto-salgado-wiedemann,vitev05,majumder05}.
\end{description}
The most simple empirically testable (and most easily theoretically computable) 
consequence of jet quenching is the suppression of inclusive high $p_T$ hadron 
spectra relative to their production in proton-proton collisions in free space. 
Since most of the energy of the fragmenting parton goes into a single {\it leading} 
hadron carrying a large fraction of the original parton energy
($\mean{z}=p_{\ensuremath{\it hadron}}/p_{\ensuremath{\it parton}}\approx$ 
0.5 -- 0.7 for $p_{\ensuremath{\it hadron}}\gtrsim$ 4 GeV/$c$ at RHIC energies); 
non-Abelian energy loss should result in a significantly suppressed 
production of high $p_T$ hadrons~\cite{gyulassy90}. The amount of suppression is 
proportional to two physical properties of the medium~\cite{bdmps,glv,wiedemann}:
\begin{description}
\item (i) the initial parton (gluon) density, $dN^g/dy$, or, 
equivalently, the transport coefficient $\hat{q} = \langle k_T^2\rangle/\lambda$ 
(which measures the average transverse momentum squared transferred to the 
projectile parton per unit path length),
\item (ii) the square of the traversed path-length, $L^2$.
\end{description}
In this contribution, experimental results on inclusive single hadron 
production at high $p_T$ in nucleus-nucleus (A+A) collisions at top CERN-SPS energies 
($\sqrt{s}\approx$ 20 GeV), intermediate RHIC energies ($\sqrt{s}$ = 62.4 GeV), 
and maximum RHIC energies ($\sqrt{s}$ = 200 GeV), are reviewed and confronted
to the theoretical predictions (i) and (ii) of non-Abelian parton energy loss 
in a dense QCD medium.\\

In the absence of any initial- or final-state medium effect, 
the total hard interaction probability in a given A+B collision 
is only due to independent parton collisions which
add incoherently. Based on individual point-like scattering and 
general QCD factorization arguments~\cite{dde_qm04},
the total hard cross-sections in A+B collisions can be written as 
the incoherent sum of all possible interactions between partons 
of nucleus A and partons of nucleus B. Namely,
\begin{equation}
E\,d\sigma_{AB\rightarrow hX}^{hard}/d^3p=A\cdot B\cdot E\,d\sigma_{pp\rightarrow hX}^{hard}/d^3p,
\label{eq:AB_scaling}
\end{equation}
where A (B) is the mass number (i.e. the number of nucleons) of nucleus A (B).
Direct experimental measurements of hard processes in nuclear collisions,
such as Drell-Yan production in Pb+Pb at CERN-SPS~\cite{na50_drellyan}, or 
prompt-$\gamma$~\cite{justin_christian,phenix_AuAugamma} and total charm yields~\cite{phenix_AuAucharm}
in Au+Au at RHIC, support such a scaling.
For a given centrality bin, Eq.~(\ref{eq:AB_scaling}) translates into 
the so-called ``$N_{coll}$ (binary) scaling'' between hard 
p+p cross-sections and A+B yields:
\begin{equation}
E\,dN_{AB\rightarrow hX}^{hard}/d^3p\,(b)=\langle T_{AB}(b)\rangle\cdot E\,d\sigma_{pp\rightarrow hX}^{hard}/d^3p,
\label{eq:TAB_scaling}
\end{equation}
where $T_{AB}(b)$ (Glauber nuclear overlap function) gives the number 
of nucleon-nucleon ($NN$) collisions in the A+B transverse overlap area 
at impact parameter $b$.
The standard method to quantify the effects of the medium in a given 
hard probe produced in a A+A reaction is given by the {\it nuclear modification factor}:
\begin{eqnarray} 
R_{AA}(p_{T},y;b)&=&\frac{\mbox{\small{``hot/dense QCD medium''}}}{\mbox{\small{``QCD vacuum''}}}\,= \nonumber \\
&=&\frac{d^2N_{AA}/dy dp_{T}}{\langle T_{AA}(b)\rangle\,\cdot\, d^2 \sigma_{pp}/dy dp_{T}},
\label{eq:R_AA}
\end{eqnarray}
which measures the deviation of A+A at $b$ from an incoherent 
superposition of $NN$ collisions, at transverse
momentum $p_T$ and rapidity $y$.

\section{A+A collisions at $\sqrtsnn\approx$ 20 GeV.}
\label{sec:sps}

\begin{figure}[htbp]
\resizebox{0.5\textwidth}{!}{
\includegraphics[height=3.cm,keepaspectratio=true]{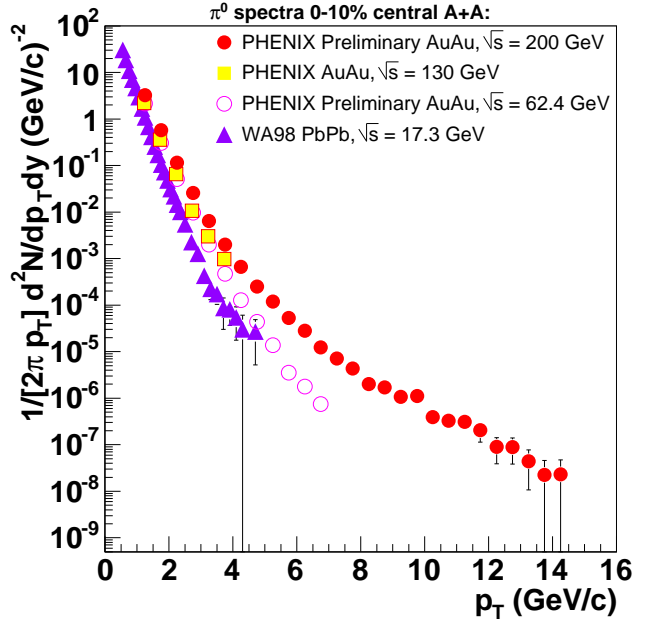}
}
\caption{High $p_T$ neutral pion spectra measured in central A+A collisions
at different center-of-mass energies: $\sqrtsnn$ = 17.3 GeV~\protect\cite{wa98},
62.4 GeV~\protect\cite{phenix_hipt_62}, 130 GeV~\protect\cite{phenix_hipt_130}, 
and 200 GeV~\protect\cite{phenix_hipt_pi0_200,phenix_hipt_pi0_eta_200}.}
\label{fig:pi0_spectra}
\end{figure}

Three nucleus-nucleus experiments at CERN-SPS measured hadron production above 
$p_T$ = 2 GeV/$c$. WA98 and CERES/NA45 measured $\pi^0$ and $\pi^\pm$ in Pb+Pb~\cite{wa98}
and Pb+Au~\cite{ceres} reactions at $\sqrt{s_{\mbox{\tiny{\it{NN}}}}}$ = 17.3 GeV
resp., whereas WA80 measured $\pi^0$ in S+Au at 
$\sqrt{s_{\mbox{\tiny{\it{NN}}}}}$ = 19.4 GeV~\cite{wa80}.
At these relatively low center-of-mass energies, the cross-sections for
hard-scattering are extremely low and the maximum transverse momenta 
measured was $p_T\approx$ 4 GeV/$c$ in the three cases\footnote{Which, yet, 
roughly corresponded to a remarkable $\sim$1/2 of 
the kinematical limit, $p_T^{\ensuremath{\it max}} = \sqrt{s}/2 \approx$ 9 GeV/$c$.}. 
Nonetheless, the power-law tail characteristic of elementary parton-parton interactions
is clearly apparent in the measured A+A spectra above $p_T\approx$ 2 GeV/$c$ 
(Fig.~\ref{fig:pi0_spectra}). Unfortunately, no baseline high $p_T$ pions 
were measured in p+p collisions at SPS at the same c.m. energies as heavy-ions, and extrapolations 
from higher-$\sqrt{s}$ data were used 
to obtain the expected $d^2\sigma_{pp}/dp_Tdy$ spectrum needed to compute the
nuclear modification factor, via Eq.~(\ref{eq:R_AA}).

\begin{figure}[htbp]
\resizebox{0.5\textwidth}{!}{
\includegraphics[height=3.3cm,width=4.4cm]{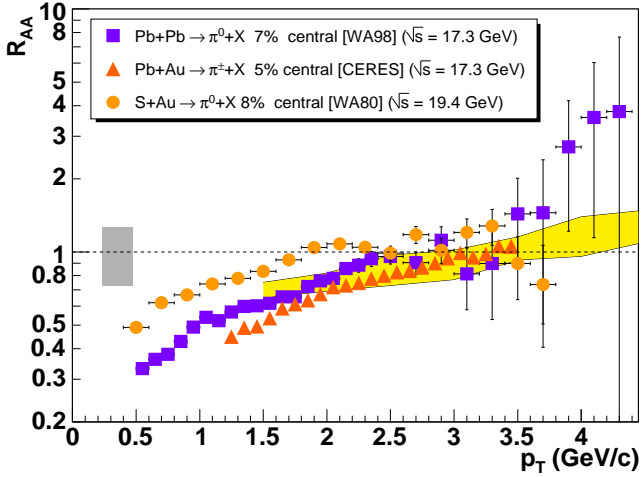}
}
\caption{Nuclear modification factors for pions produced at CERN-SPS 
in central Pb+Pb~\protect\cite{wa98}, Pb+Au~\protect\cite{ceres}, 
and S+Au~\protect\cite{wa80} at $\sqrt{s_{\mbox{\tiny{\it{NN}}}}}\approx$ 20 GeV 
obtained using the p+p parametrization proposed in~\protect\cite{dde_hipt_sps},
compared to a theoretical prediction
~\protect\cite{vitev_gyulassy} of 
final-state parton energy loss in a system with initial gluon 
densities $dN^g/dy=$ 400 -- 600. The shaded band at $R_{AA}$ = 1 
represents the overall fractional uncertainty of the data
(CERES data~\protect\cite{ceres} have an extra 
uncertainty of $\pm$15\% not shown).}
\label{fig:RAA_centSPS_vs_vitev}
\end{figure}

In ~\cite{dde_hipt_sps}, we compared several proposed 
p+p $\rightarrow \pi^{0,\pm}+X$ parametrizations, to the existing 
data in the range $\sqrt{s}\approx$ 16 -- 20 GeV, and found that the 
parametrization of Blattnig {\it et. al}~\cite{blatt} reproduced reasonably well, 
within $\sim$25\%, the shape and magnitude of the experimental pion differential 
cross-sections below $p_T\approx$ 4 GeV/$c$. Using this p+p reference, we
obtained the nuclear modification factors for central A+A collisions at SPS
shown in Fig.~\ref{fig:RAA_centSPS_vs_vitev}. 
Hadron production below $p_T\approx$ 1 GeV/$c$ falls, as expected, below $R_{AA}$ = 1 
(the assumption of independent point-like scattering does not hold for soft processes),
but high-$p_T$ hadroproduction 
is, within errors, consistent with scaling with the number of $NN$ collisions.
Such a result is at variance with the factor of $\sim 2$ {\it enhancement} 
observed in high $p_T$ pion production in {\it peripheral} Pb+Pb reactions 
at the same energies~\cite{wa98,dde_hipt_sps} attributed to 
initial-state $p_T$ broadening as observed in fixed-target $p+A$ reactions
(``Cronin effect'')~\cite{cronin}.
This result points to the existence of an attenuating mechanism that 
reduces the underlying Cronin enhancement from $R_{AA}\gtrsim$ 2
down to values consistent with $R_{AA}\approx$ 1 in central A+A. Indeed, 
theoretical predictions of high $p_T$ $\pi^0$ production in central Pb+Pb 
at SPS including Cronin broadening, nuclear-modified parton distribution 
functions (PDF), and final-state partonic energy loss in an 
expanding system with initial effective\footnote{Note that at SPS energies, the
initial medium is probably more ``quarkonic'' than ``gluonic''.} gluon densities 
$dN^g/dy=$ 400 -- 600~\cite{vitev_gyulassy} reproduce well the observed nuclear 
modification factor (yellow band in Fig.~\ref{fig:RAA_centSPS_vs_vitev}).

The interesting conclusion that a moderate amount of jet quenching is already 
present in the most central heavy-ion reactions at SPS would require, however, 
a direct (and accurate) measurement of the high $p_T$ p+p pion spectrum
at $\sqrt{s}$ = 17.3 GeV. Unfortunately, the default minimum collision energy 
at RHIC in the proton-proton mode is $\sqrt{s}\approx$ 48 GeV~\cite{dde_hq04}, 
making it difficult to directly compare at RHIC high $p_T$ hadroproduction in 
A+A and p+p collisions at center-of-mass energies comparable to SPS.

\section{A+A collisions at $\sqrtsnn$ = 62.4 GeV.}
\label{sec:rhic62}

The study of the excitation function of high $p_T$ hadron suppression between 
top SPS and top RHIC energies was the main motivation behind the dedicated 
Au+Au run at RHIC intermediate energies ($\sqrt{s_{\mbox{\tiny{\it{NN}}}}}$ = 62.4 GeV)
carried out in April 2004. PHENIX measured neutral pions in the range 
$p_T$ = 1 -- 7 GeV/$c$~\cite{phenix_hipt_62}. 
PHOBOS~\cite{phobos_hipt_62} and STAR~\cite{star_hipt_62} measured inclusive 
charged hadrons up to $p_T\approx$ 4.5 GeV/$c$ and 12 GeV/$c$ respectively. 
However, as in the SPS case, no concurrent p+p reference measurement was performed
at $\sqrt{s}$ = 62.4 GeV, and the corresponding Au+Au nuclear modifications factors
were determined using p+p $\rightarrow$ $h^\pm,\pi^0 + X$ 
differential cross-sections measured at the top CERN-ISR energies 
($\sqrt{s}$ = 62 $-$ 63 GeV) in the 70s and 80s. As discussed in~\cite{dde_hq04},
the existing large inconsistencies (up to a factor of $\sim$3) among the different
ISR $\pi^0$ data sets can be greatly reduced by removing the direct-$\gamma$
and $\eta$ contaminations not subtracted from the original ``unresolved'' 
$\pi^0$ measurements. By doing so, one can obtain an averaged 
$d^2\sigma_{pp\rightarrow\pi^0}/dp_Tdy$ reference spectrum at 
$\sqrt{s}$ = 62.4 GeV with uncertainties $\pm$25\% 
for the $R_{AA}$ denominator. 

\begin{figure}[htbp]
\resizebox{0.5\textwidth}{!}{
\includegraphics[height=5.5cm]{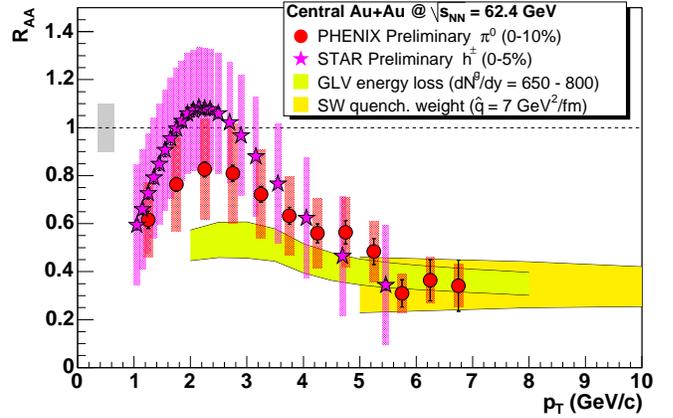}
}
\caption{Preliminary PHENIX and STAR nuclear modification factors, $R_{AA}(p_T)$, 
for $\pi^0$~\protect\cite{phenix_hipt_62} and $h^\pm$~\protect\cite{star_hipt_62} 
obtained in central Au+Au at $\sqrtsnn$ = 62.4 GeV using the p+p $\rightarrow \pi^{0},h^\pm+X$ 
references discussed in~\protect\cite{dde_hq04,star_hipt_62}. The data are
compared to two theoretical predictions for parton energy loss in a dense medium 
with initial gluon density $dN^g/dy=$ 650 -- 800~\protect\cite{vitev_62.4} or, 
equivalently, transport coefficient $\mean{\hat{q}}$ = 7 GeV$^2$/fm~\cite{dainese04,eskola04}. 
The error ``bands'' around each data point indicate the $\sim$25\% systematic uncertainty 
of the ISR p+p baseline spectra at $\sqrt{s}$ = 62.4 GeV.
The shaded band at $R_{AA}$ = 1 represents the overall fractional uncertainty of the data
(absolute normalization of Au+Au spectra and $\mean{T_{AA}}$ uncertainties).}
\label{fig:R_AA_62.4}
\end{figure}

Figure~\ref{fig:R_AA_62.4} shows the PHENIX and STAR preliminary nuclear 
modification factors for high $p_T$ $\pi^0$~\cite{phenix_hipt_62} and 
$h^\pm$~\cite{star_hipt_62} in central Au+Au collisions at 
$\sqrt{s_{\mbox{\tiny{\it{NN}}}}}$ = 62.4 GeV obtained using the p+p 
references discussed in~\cite{dde_hq04,star_hipt_62}.
Above $p_T\approx$ 5 GeV/$c$, there are $\sim$3 times less produced pions and 
inclusive hadrons than expected from point-like scaling of the p+p cross-sections.
The magnitude of the suppression can be reproduced by models that include parton 
energy loss in a system with initial gluon densities 
$dN^g/dy=$ 650 -- 800~\cite{vitev_62.4,adil_gyulassy04,wang04} (GLV formalism~\cite{glv}, 
green band) or transport coefficients $\mean{\hat{q}}\approx$ 7 GeV$^2$/fm~\cite{dainese04,eskola04} 
(Salgado-Wiedemann quenching weights~\cite{wiedemann}, yellow band).
The range $p_T\approx$ 1 -- 5 GeV/$c$ shows, however, a significant rise and fall 
of the Au+Au spectra compared to the scaled p+p reference\footnote{An effect which 
is larger for the charged hadrons than for $\pi^0$ due to the observed enhanced 
(anti)proton production~\cite{phenix_hipt_ppbar}.}. Although part of this effect can 
be certainly attributed to the expected collective radial flow and 
Cronin~\cite{accardi05}-recombination~\cite{greco05} enhancement$^3$ at 62.4 GeV, 
there are also $p_T$-dependent uncertainties related to the relatively poorly 
known shape of the ISR p+p spectra in this intermediate $p_T$ range~\cite{dde_hq04}.
Clearly, a dedicated RHIC proton-proton run at this collision energy would help to 
reduce these uncertainties and better constraint the theoretical predictions
of parton energy loss models.

\section{A+A collisions at $\sqrtsnn$ = 200 GeV.}
\label{rhic200}

\begin{figure}[htbp]
\resizebox{0.5\textwidth}{!}{
\includegraphics[height=3.5cm,width=5.5cm]{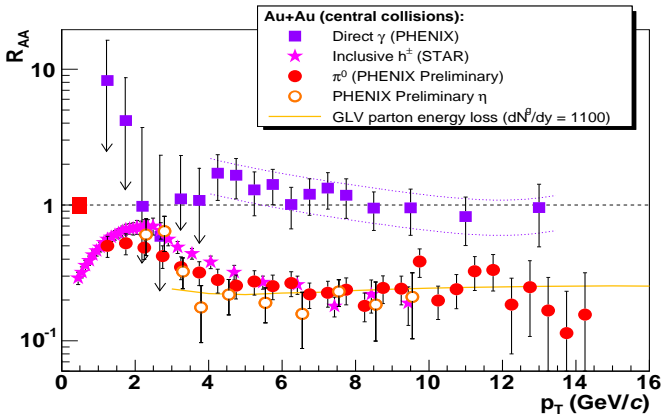}}
\caption{$R_{AA}(p_T)$ measured in central Au+Au at 200 GeV for: 
direct photons~\protect\cite{justin_christian,phenix_AuAugamma},
inclusive charged hadrons~\protect\cite{star_hipt_200},
$\pi^0$~\protect\cite{phenix_hipt_pi0_eta_200}, and 
$\eta$~\protect\cite{phenix_hipt_pi0_eta_200}
compared to theoretical predictions for parton energy loss in a dense medium with 
$dN^g/dy=$ 1100~\protect\cite{vitev_gyulassy}. The shaded band at $R_{AA}$ = 1 
represents the overall fractional uncertainty of the data (absolute normalization 
of spectra and $\mean{T_{AA}}$ uncertainties). The baseline p+p reference of the 
direct $\gamma$ Au+Au data is a NLO calculation whose uncertainties are indicated by 
the dotted lines around the points~\protect\cite{phenix_AuAugamma}.}
\label{fig:R_AA_RHIC_200}
\end{figure}

Undoubtedly, one of the most significant results from the first 4 years
of operation at RHIC is the large high $p_T$ hadron suppression observed 
in central Au+Au reactions at $\sqrtsnn$ = 200 GeV. Above 
$p_T\approx$ 5 GeV/$c$, pions~\cite{phenix_hipt_pi0_200}, 
eta mesons~\cite{phenix_hipt_pi0_eta_200}, and inclusive hadrons 
($h^\pm$)~\cite{star_hipt_200,phenix_hipt_200} show all a ``universal'' factor of 
$\sim$5 suppression compared to the corresponding $T_{AA}$-scaled proton-proton yields.
Such a deficit is {\it not} observed for direct photons~\cite{justin_christian,phenix_AuAugamma} (Fig.~\ref{fig:R_AA_RHIC_200}).
The observed hadron suppression remains constant as a function of $p_T$ up to the highest transverse 
momenta measured so far ($p_T\approx$ 14 GeV/$c$ for $\pi^0$~\cite{phenix_hipt_pi0_eta_200}, 
see Fig.~\ref{fig:R_AA_RHIC_200}) in agreement with the parton energy loss model
predictions. In the first theoretical predictions~\cite{bdmps}, approximations of 
the underlying Landau-Pomeranchuk-Migdal (LPM) interference effect in gluon 
bremsstrahlung resulted in a logarithmic dependence of the quenching factor on 
the parton energy and, therefore, in a $R_{AA}(p_T)$ that (slowly) increased
for increasing hadron $p_T$'s in apparent contradiction with the data. 
However, (i) the use of a realistic energy distribution of the emitted gluons 
(rather than the mean value)~\cite{jeon_moore}, (ii) finite kinematic constraints 
(in the energy loss and in-medium path length), and/or depleted nuclear PDFs 
(in the EMC region for Bjorken $x\approx \frac{2p_T/\mean{z}}{\sqrtsnn}\gtrsim 0.2$ 
values corresponding to $p_T\gtrsim$ 12 GeV/$c$)~\cite{vitev_gyulassy} and, 
(iii) the increasing power-law (local) exponent of the parton spectra with 
$p_T$~\cite{eskola04}, all explain the effectively constant $R_{AA}$ evolution
as a function of transverse momentum.

\begin{figure}[htbp]
\epsfig{file=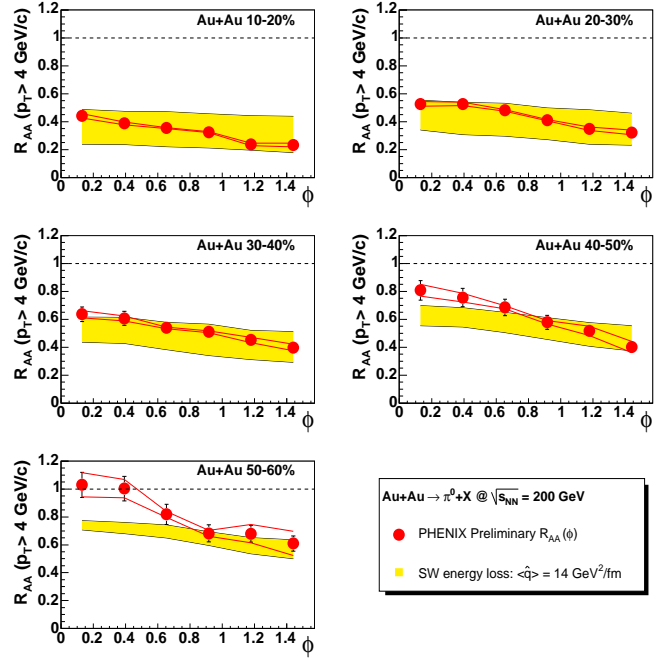,height=9.cm,width=9.cm}
\caption{Preliminary PHENIX nuclear modification factor, $R_{AA}(\phi)$, 
for $\pi^0$ production above $p_T$ = 4 GeV/$c$ as a function of the azimuthal 
angle $\phi$ with respect to the reaction plane in 5 centrality classes of Au+Au 
at $\sqrtsnn$ = 200 GeV~\protect\cite{cole}; compared to parton energy loss 
calculations~\protect\cite{dainese04} for an azimuthally asymmetric system
with average transport coefficient $\mean{\hat{q}}\approx$ 14 GeV$^2$/fm 
(yellow band, encompassing the limits of two different prescriptions 
for the sampling of the energy loss). The lines around the experimental data 
show the uncertainties in the reaction plane and $R_{AA}$ determination.}
\label{fig:RAA_vs_phi}
\end{figure}

A robust prediction of non-Abelian parton energy loss calculations is the expected 
$\propto L^2$ dependence of the average energy loss as a function of the in-medium path 
length~\cite{bdmps}. Such a behaviour predicted for a {\it static} QCD medium, 
turns into an effective $\propto L$-dependence in an expanding QGP~\cite{glv}. 
An interesting way to experimentally test the $L$ dependence of the energy 
loss is by exploiting the spatial azimuthal asymmetry of the system 
produced in non-central nuclear collisions. Indeed, due to the characteristic 
almond-like shape of the overlapping matter produced in A+A reactions with 
finite impact parameter, partons traversing the produced medium along the direction 
perpendicular to the reaction plane (``out-of-plane'') will comparatively go 
through more matter than those going parallel to it (``in-plane''), and therefore
are expected to lose more energy. In general, the total path length along a 
given azimuthal angle $\phi$ with respect to the reaction plane is 
$L(\phi)\approx 1 - (\varepsilon/2)\,cos(2\phi)$~\cite{cole}, where $\varepsilon$ is the
eccentricity of the system. By looking at the suppression pattern along
different $\phi$ trajectories one can test the $L$ dependence of the energy loss. 
PHENIX~\cite{cole} has recently presented nuclear modification factors for high 
$p_T$ $\pi^0$ in Au+Au collisions at $\sqrtsnn$ = 200 GeV binned in $\phi$ 
angle with respect to the reaction plane (determined with the Beam-Beam-Counters 
at high rapidities). The resulting $R_{AA}(\phi)$ curves (Fig.~\ref{fig:RAA_vs_phi}) 
show clearly a factor of $\sim$2 more suppression out-of-plane ($\phi$ = $\pi$/2) 
than in-plane ($\phi$ = 0) for all the centralities (eccentricities) considered. 
Theoretical calculations of parton energy loss (based in the quenching weights 
formalism ~\cite{wiedemann}) in an azimuthally asymmetric medium~\cite{dainese04} 
predict a significantly smaller difference between the suppression patterns for
partons emitted at $\phi$ = 0 and $\phi$ = $\pi$/2 (yellow bands in 
Fig.~\ref{fig:RAA_vs_phi}). The discrepancy model--data is stronger for more 
peripheral centralities (with correspondingly larger eccentricities) and 
challenges the underlying in-medium path-length dependence of non-Abelian 
parton energy loss (a detailed discussion of this observation can be found 
in~\cite{cole}) and/or points out the necessity of an additional source
of azimuthal anisotropy in pion production at high $p_T$. Likely, collective 
elliptic flow is responsible for the extra boost of in-plane pions (even at
$p_T$ values above 4 GeV/$c$), though this should be confirmed quantitatively.

\section{QCD medium properties via ``jet tomography''}
\label{sec:medium_prop}

\begin{figure}[htbp]
\epsfig{file=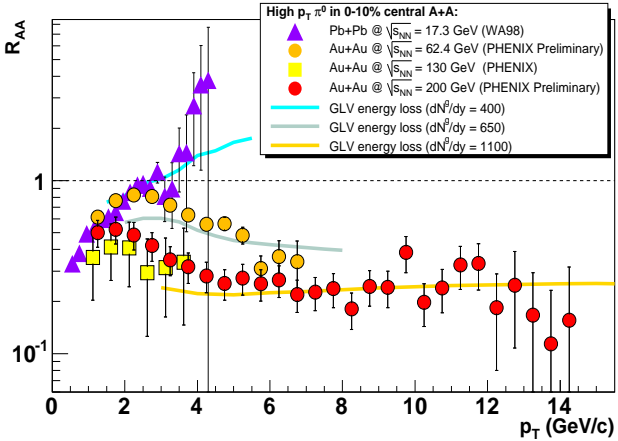,height=6.0cm,width=8.5cm}
\caption{Nuclear modification factor, $R_{AA}(p_T)$, for high $p_T$ pion production
in central nucleus-nucleus reactions in Pb+Pb~\protect\cite{wa98} at 
$\sqrt{s_{\mbox{\tiny{\it{NN}}}}}$ = 17.3 GeV; and 
Au+Au at $\sqrtsnn$ = 62.4 GeV~\protect\cite{phenix_hipt_62}, 
$\sqrtsnn$ = 130 GeV~\protect\cite{phenix_hipt_130}, and 
$\sqrtsnn$ = 200 GeV~\protect\cite{phenix_hipt_pi0_200}; compared to GLV parton 
energy loss calculations~\protect\cite{vitev_gyulassy,vitev_62.4} for 
different initial gluon densities ($dN^g/dy$ = 400, 650 and 1100).
Absolute normalization errors of the data, $\mathcal{O}$(10\%--25\%), 
are not shown.}
\label{fig:RAA_compilation}
\end{figure}

Fig.~\ref{fig:RAA_compilation} compiles all the available $R_{AA}(p_T)$ for 
high $p_T$ (leading) neutral pions measured in central A+A collisions in the range 
$\sqrtsnn\approx$ 20 -- 200 GeV. The experimental suppression factors can
be well reproduced by parton energy loss calculations that assume the
formation of strongly interacting systems with initial gluon densities 
per unit rapidity in the range $dN^g/dy\approx$ 400 -- 1200~\cite{vitev_gyulassy,vitev_62.4} 
or, equivalently~\cite{wiedemann}, with time-averaged transport coefficients 
$\mean{\hat{q}}\approx$ 3.5 -- 15 GeV$^2$/fm~\cite{dainese04} (see Table~\ref{tab:1}). 

\begin{table}[htbp]
\caption{Effective initial gluon densities, $dN^g/dy$,~\protect\cite{vitev_gyulassy,vitev_62.4}
and time-averaged transport coefficients $\mean{\hat{q}}$~\protect\cite{dainese04},
for the strongly interacting media produced in central A+A collisions at 
four different center-of-mass energies obtained from parton energy loss calculations 
reproducing the observed high $p_T$ leading $\pi^0$ suppression at each collision 
energy (``jet tomography''). The measured charged particle multiplicity densities 
at mid-rapidity~\protect\cite{ppg019}, $dN_{ch}^{exp}/d\eta$, are also quoted for 
each $\sqrtsnn$.}
\label{tab:1}
\begin{tabular}{lcccc}
\hline\noalign{\smallskip}
 & $\sqrtsnn$ & $dN^g/dy$ & $\mean{\hat{q}}$ & $dN_{ch}^{exp}/d\eta$  \\
 &  (GeV) &  & (GeV$^2$/fm) &  \\
\noalign{\smallskip}\hline\noalign{\smallskip}
SPS  &  17.3 & 400   & 3.5 & 312 $\pm$ 21 \\
RHIC &  62.4 & 650   & 7.  & 475 $\pm$ 33 \\
RHIC &  130. & $\sim$900    & $\sim$11  & 602 $\pm$ 28 \\
RHIC &  200. & 1100 	    & 14. -- 15. & 687 $\pm$ 37 \\
\noalign{\smallskip}\hline
\end{tabular}
\end{table}

For each collision energy, the derived values for the initial rapidity 
density, $dN^g/dy$, and transport coefficient $\mean{\hat{q}}$, 
are consistent with each other and with the final particle density 
measured in the reactions. Indeed, assuming an isentropic expansion process, 
all the hadrons produced at mid-rapidity in a A+A collision come directly 
from the original gluons released in the initial phase of the reaction\footnote{We 
use here: $N_{tot}/N_{ch}$ = 3/2, and $|d\eta/dy|\approx$ 1.2.}:
\begin{equation}
\frac{dN^g}{dy}\approx\frac{N_{tot}}{N_{ch}}\,\left|\frac{d\eta}{dy}\right|\,\frac{dN_{ch}}{d\eta}
\approx 1.8\cdot\frac{dN_{ch}}{d\eta}.
\label{eq:dNgdy}
\end{equation}
This relation is relatively well fulfilled by the data as can be seen by comparing
columns third and fifth of Table~\ref{tab:1}. The time-dependent transport 
coefficient scales with the energy density of the medium\footnote{Rather than a 
thermodynamical variable, $\hat{q}$ is actually a {\it dynamical} quantity resulting 
from the product of the time-dependent density of scattering centers times the 
strength of each single elastic scattering~\cite{wiedemann}.} ($\varepsilon$ in GeV/fm$^3$) 
like~\cite{eskola04}:
\begin{equation}
\hat{q}(\tau)\approx 8\cdot\varepsilon^{3/4}(\tau).
\label{eq:qhat}
\end{equation}
Since, for an ideal QGP (with 2+1 flavors, i.e. 
degeneracy $g\approx$ 42), 
the particle ($\rho \approx 4.7\cdot (T/\hbar c)^{3}$) and energy ($\varepsilon \approx 14\cdot T^4/(\hbar c)^{3}$) 
densities~\cite{wong} are related via 
$\rho \approx 0.66\cdot(\varepsilon/\hbar c)^{3/4}\approx 2.3\cdot\varepsilon^{3/4}$, 
one can express Eq. (\ref{eq:qhat}) as
\begin{equation}
\hat{q}(\tau)\approx 3.5\cdot\rho(\tau) = 3.5\cdot\rho_0\left(\frac{\tau_0}{\tau}\right) = 
3.5\cdot\frac{dN^{g}}{dV}\left(\frac{\tau_0}{\tau}\right) \, ,
\label{eq:qhat2}
\end{equation}
where for the second equality we have assumed a 1-dim. Bjorken expansion.
In this scenario, since the medium expands boost-invariantly in the longitudinal 
direction, we can further write $dV = A_T\,\tau_0\,dy$ where $A_T$ is the transverse 
area of the system, and therefore
\begin{equation}
\hat{q}(\tau)\approx \frac{3.5}{A_T}\cdot\frac{dN^{g}}{dy}\cdot\frac{1}{\tau} .
\label{eq:qhat3}
\end{equation}
According to \cite{wiedemann}, the relation between the time-averaged $\hat{q}(\tau)$ 
in an expanding medium and that of a fixed static medium is (taking $\tau_f\gg\tau_0$):
\begin{equation}
\mean{\hat{q}(\tau)} = \frac{2}{L_{\ensuremath{\it eff}}^2}\int_{\tau_{0}}^{\tau_0+L_{\ensuremath{\it eff}}}\!\!\!\!\!\!(\tau - \tau_0)\,\hat{q}(\tau)\, d\tau
\approx \frac{2}{L_{\ensuremath{\it eff}}}\,\frac{3.5}{A_T}\,\frac{dN^{g}}{dy}
\label{eq:qhat4}
\end{equation}
where $L_{\ensuremath{\it eff}}$ is the effective length traversed by the parton
in the medium. Despite the simplifying assumptions used, this approximate relation between 
the medium transport coefficient and the original gluon rapidity density is relatively 
well fulfilled by the data too (Table~\ref{tab:1}). E.g. by taking 
$L_{\ensuremath{\it eff}}\approx$ 4 fm and 
$\langle A_T\rangle\approx$ 125 fm$^2$ for 0-10\% central Au+Au we get
\begin{equation}
\langle \hat{q}(\tau)\rangle \approx 0.014\cdot\frac{dN^{g}}{dy}.
\label{eq:qhat_rhic}
\end{equation}

\section{Excitation function of high $p_T$ leading hadron suppression}
\label{sec:exc_funct}

\begin{figure}[htbp]
\resizebox{0.45\textwidth}{!}{%
\includegraphics[height=9.5cm,width=5.cm,angle=-90]{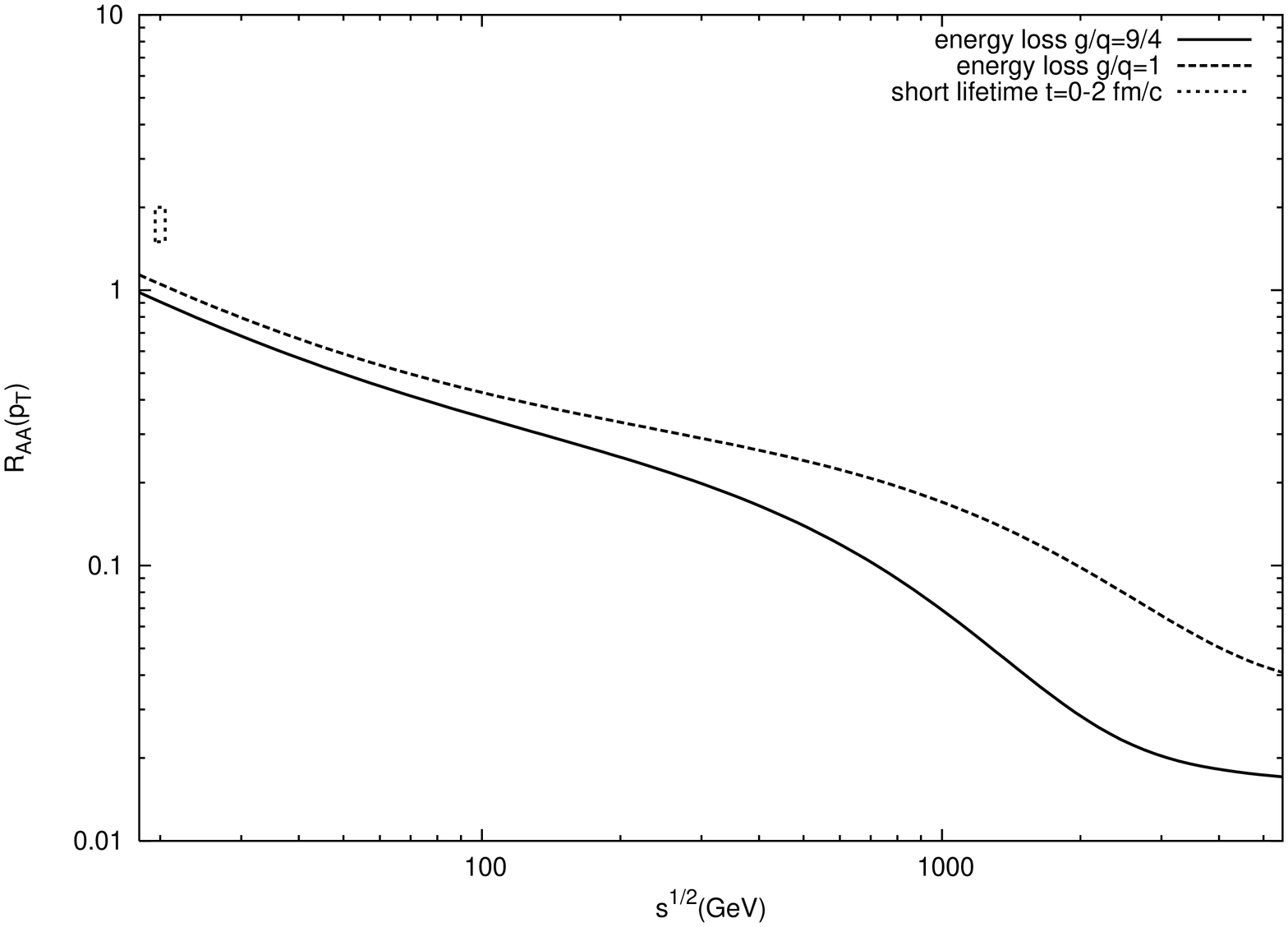}}
\resizebox{0.44\textwidth}{!}{%
\includegraphics[height=4.5cm,width=8.cm]{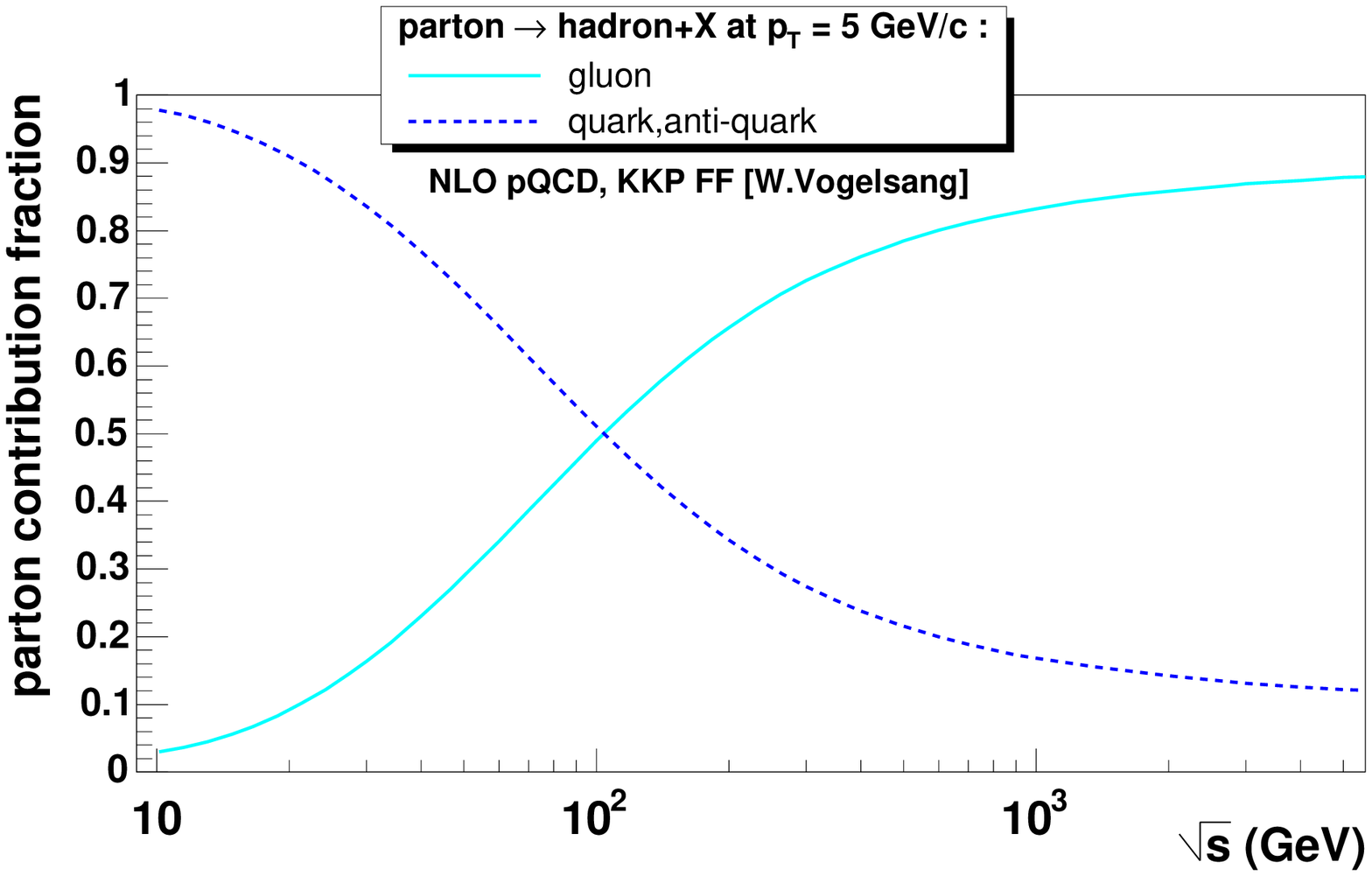}}
\caption{Top (Fig. taken from~\protect\cite{wang05}): $R_{AA}$ for neutral pions 
as function of collision energy at fixed $p_T=6$ GeV/$c$ in 10\% most central AuAu 
collisions for non-Abelian (lower curve) and ``non-QCD'' (upper curve) energy loss patterns. 
Bottom: Relative proportion of quarks and gluons fragmenting into a hadron 
at fixed $p_T$ = 5 GeV/$c$ in p+p collisions in the range 
$\sqrt{s}$ = 10 -- 5500 GeV as given by NLO pQCD~\protect\cite{vogelsang}.}
\label{fig:q_g_RAA_fraction}
\end{figure}

Based on rather general grounds, the total amount of leading hadron suppression 
at a fixed (large) $p_T$ in central A+A collisions should depend on the 
collision energy only via two $\sqrt{s}$-dependent factors:
\begin{description}
\item (i) the initial parton density of the produced system, and 
\item (ii) the relative fraction of quarks and gluons fragmenting into the 
hadron at the $p_T$ value in question.
\end{description}
Indeed, on the one hand, since $\Delta E_{\ensuremath{\it loss}}\propto dN^g/dy \propto dN_{ch}/d\eta$,
and since the total particle multiplicity produced at mid-rapidity in A+A 
collisions is observed to follow the approximate scaling~\cite{ppg019}:
\begin{equation}
dN_{ch}/d\eta \approx 0.75\cdot(N_{part}/2)\cdot \ln(\sqrtsnn/1.5),
\label{eq:dNchdy}
\end{equation}
one expects the amount of suppression to increase accordingly with $\sqrtsnn$
(note, however, that the true evolution of the suppression will be faster than 
that given by Eq.~(\ref{eq:dNchdy}) since, for increasing energies {\it both} 
$dN^g/dy$ and the corresponding lifetime of the quenching medium are larger). 
On the other hand, the probability for a gluon to lose energy is a factor 
$C_A/C_F$ = 9/4 larger than the probability for a quark\footnote{In QCD, 
the relative strengths of the three distinct quark and gluon vertices,  
$\alpha_s C_F$ for $q \rightarrow qg$,  $\alpha_s C_A$ for $g \rightarrow gg$, 
and $\alpha_s T_F$ for $g \rightarrow q{\bar q}$ are completely determined 
by the structure of the gauge group (Casimir factors) describing the strong force. 
For $SU(N_c)$ where $N_c$ is the number of colors, $C_A = N_c$, $C_F = (N^2-1)/2N_C$
and $T_F$ = 1/2. The probability for a gluon (quark) to radiate a 
gluon is proportional to the color factor $C_A$ = 3 ($C_F$ = 4/3). In the asymptotic limit, 
and neglecting the splitting of gluons to quark-antiquark pairs (proportional 
to the smaller color factor $T_R$ = 1/2), the average number of gluons radiated by 
a gluon is, therefore, a factor $C_A/C_F$ = 9/4 higher than the number of gluons radiated 
by a quark~\cite{ellis96}.}, and the relative fraction of hard scattered quarks and gluons 
(going through the medium and) fragmenting into a hadron at a fixed $p_T$ 
varies with $\sqrtsnn$  
in a proportion given by a tradeoff between (i) the relative density of quarks 
and gluons at the corresponding Bjorken $x = 2p_T/\sqrt{s}$, and (ii) the relative
fragmentation ``hardness'' of quarks and gluons at the corresponding $z$
value. A full NLO calculation~\cite{vogelsang} gives the results shown in 
Fig.~\ref{fig:q_g_RAA_fraction} (bottom). 

In reference~\cite{wang05}, Wang\&Wang presented a pQCD calculation of
the expected $\sqrtsnn$-dependence of the nuclear modification factor 
for high-$p_T$ $\pi^0$ production in central Au+Au collisions due to 
parton energy loss in the produced (2-D expanding) QGP. The resulting
curve is shown in Fig.~\ref{fig:q_g_RAA_fraction} (top). The amount
of suppression (for a 6 GeV/$c$ leading hadron) increases monotonically 
with $\sqrtsnn$ due to the growing initial parton density, QGP lifetime,
and gluonic nature of the quenched parton, and seems to saturate at 
$R_{AA}\approx$ 0.02 for c.m. energies above $\sim$3 TeV. The existence 
of a maximum amount of suppression is due to ``irreducible'' particle 
production from the outer corona of the medium, which remains unsuppressed 
even for extreme energy densities~\cite{eskola04}.

\begin{figure}[htbp]
\epsfig{file=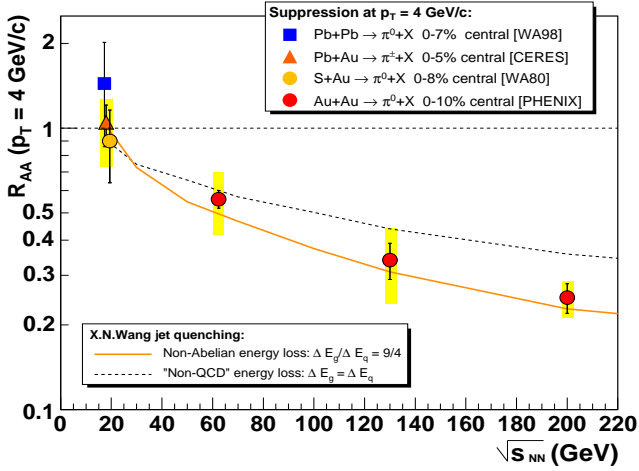,height=6.2cm,width=8.50cm}
\caption{Excitation function of the nuclear modification factor, $R_{AA}(\sqrtsnn)$, 
for $\pi^0$ production in central A+A reactions at a fixed $p_T$ = 4 GeV/$c$ value, 
compared to predictions of a jet-quenching model with canonical non-Abelian 
(solid line) and ``non-QCD'' (dashed line) energy losses~\protect\cite{wang05}.
The shaded band around each data point represents the absolute systematic errors 
(absolute normalization of A+A and p+p spectra and nuclear overlap, $\mean{T_{AA}}$, 
uncertainties)}
\label{fig:last}
\end{figure}

In order to test the effect of the radiative QCD energy loss, they compared 
the expected non-Abelian prescription (in which gluons lose 
$\Delta E_q/\Delta E_g$~=~9/4 times more energy than quarks) to an 
arbitrary ``non-QCD'' recipe in which quarks and gluons lose 
the same amount of energy ($\Delta E_q = \Delta E_g$). 
For a fixed hadron $p_T$ value, say $p_T\approx$ 6 GeV/$c$, the 
total suppression factors from non-Abelian and non-QCD energy losses
are relatively similar below $\sqrtsnn\approx$ 100 GeV, since quarks
are the dominant parton fragmenting into a high $p_T$ hadron 
(Fig.~\ref{fig:q_g_RAA_fraction}, bottom). Above $\sqrtsnn\approx$ 100 GeV, 
gluons take over as the dominant parent parton of hadrons with 
$p_T\approx$ 6 GeV/$c$ and, consequently, the $R_{AA}$ values drop
faster in the canonical non-Abelian scenario.
The experimental excitation function of high $p_T$ $\pi^0$ suppression 
in central A+A collisions supports the expected QCD radiative
energy loss behaviour as demonstrated in Fig.~\ref{fig:last}.

\section*{Summary}
Experimental results on high $p_T$ leading hadron production in nucleus-nucleus
reactions at center-of-mass energies $\sqrtsnn\approx$ 20 -- 200 GeV have 
been discussed with an emphasis on the observed suppression of the per-nucleon yields
relative to p+p collisions at the same $\sqrt{s}$. The amount of suppression 
steadily increases from CERN-SPS energies (Pb+Pb at $\sqrtsnn\approx$ 20 GeV) 
reaching a maximum quenching factor of $\sim$5 at the highest RHIC energies 
(Au+Au at $\sqrtsnn$ = 200 GeV) due to the increased initial parton density, 
lifetime of the dissipative medium, and gluonic nature of the parent 
fragmenting parton. The $p_T$ and $\sqrtsnn$ dependences of the measured nuclear 
modification factors are in agreement with theoretical calculations of final-state 
non-Abelian energy loss in a dense QCD medium. The observed dependence 
of the suppression factors on the reaction plane orientation is, however, 
significantly stronger than the one expected from the $\propto L$ in-medium 
path-length dependence predicted by jet quenching models, and points to an 
additional source of azimuthal anisotropy (likely collective elliptic flow) 
in hadron production at $p_T$ values above 4 GeV/$c$.

\section*{Acknowledgments}
I would like to thank Andrea~Dainese, Carlos~Salgado, Ivan~Vitev, Werner~Vogelsang, 
and Xian-Nian~Wang for providing different theoretical results confronted with 
the experimental data presented in this paper as well as for useful discussions.


\end{document}